\documentclass[12pt,a4]{article}
\usepackage{amsfonts}
\newcommand{\text}{\rm}

\usepackage[dvips]{epsfig}
\usepackage[dvips]{graphics}

\begin{document}

\title{\textbf{Perturbing Topological Field Theories}}
\author{\textbf{Vitor E.R. Lemes, Cesar A. Linhares, } \and \textbf{Silvio P.
Sorella, Luiz Claudio Q.Vilar} \\
Instituto de F\'{\i}sica, \\
Universidade do Estado do Rio de Janeiro\\
Rua S\~{a}o Francisco Xavier, 524\\
20550-013, Maracan\~{a}, Rio de Janeiro \vspace{2mm}\\
and\vspace{2mm} \and \textbf{D. G. G. Sasaki} \\
Centro Brasileiro\textbf{\ }de Pesquisas F\'{\i}sicas \\
Rua Xavier Sigaud 150, 22290-180 Urca \\
Rio de Janeiro, Brazil\vspace{2mm} \and \textbf{UERJ-DFT/01/99}\vspace{2mm}%
\newline \and \textbf{PACS: 11.10.Gh}}
\maketitle

\begin{abstract}
The abelian Chern--Simons theory is perturbed by introducing local
gauge-invariant interaction terms depending on the curvature. The
computation of the correlation function $\left\langle \oint_{\gamma
_1}dx^\mu A_\mu \oint_{\gamma _2}dy^\nu A_\nu \right\rangle $ for two
smooth closed
nonintersecting  curves $\gamma _1$, $\gamma _2$ is reported up
to four loops and is shown to be unaffected by radiative corrections. This
result ensures the stability of the linking number of $\gamma _1$ and $%
\gamma _2$ with respect to the local perturbations which may be added to the
Chern--Simons action.

\setcounter{page}{0}\thispagestyle{empty}
\end{abstract}

\vfill\newpage\ \makeatother

\section{\ Introduction\-}

Since their introduction \cite{w1}, topological field theories have been
responsible for many applications \cite{bbrt} and are object of continuous
investigations. Nowadays they represent an important chapter of quantum
field theory. The original motivation was related to the possibility of
describing topological invariants by means of standard field-theory
techniques \cite{w1,gu}.

In order to give an idea of this framework, let us briefly present here the
field-theory characterization of one of the most simple and familiar
topological invariants, namely, the linking number $\chi (\gamma _1,\gamma
_2)$ of two nonintersecting smooth closed oriented curves in $\mathcal{R}%
^3\; $\cite{gu,tn}:

\begin{center}
\begin{figure}[h]
\setlength{\unitlength}{1mm}
\hspace{2cm}\scalebox{0.75}{\includegraphics*{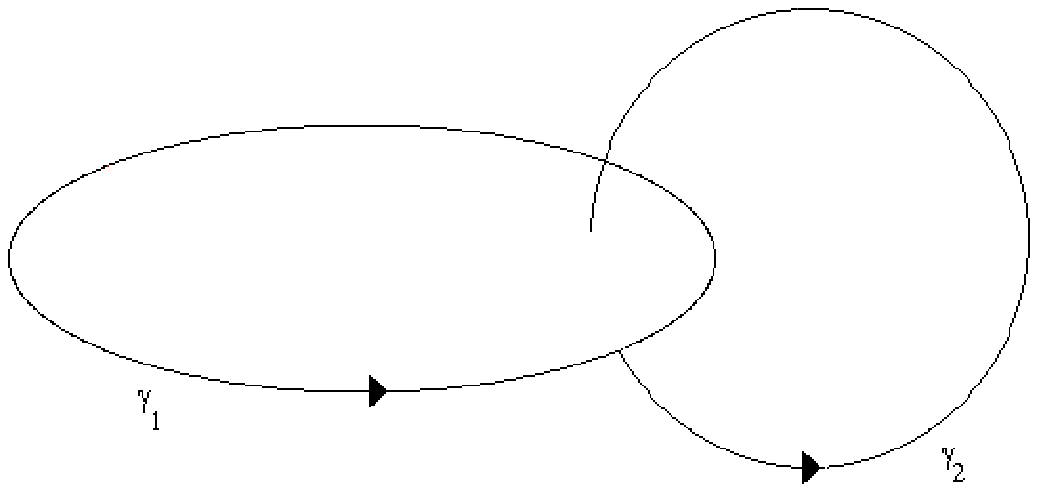}}
\caption{}
\end{figure}
\end{center}


As is well known, the linking number $\chi (\gamma _1,\gamma _2)$ is an
integer which counts the number of times that one curve winds around the
other. It is independent from the shape of the curves and can be represented
by the Gauss integral
\begin{equation}
\chi (\gamma _1,\gamma _2)=\frac 1{4\pi }\oint_{\gamma _1}dx^\mu
\oint_{\gamma _2}dy^\nu \varepsilon _{\mu \nu \rho }\frac{(x-y)^\rho }{%
\left| x-y\right| ^3}\;.  \label{g-int}
\end{equation}
Expression (\ref{g-int}) is in fact easily seen to be an integer by use of
the Stokes' theorem \cite{gu,tn}. Taking a field theory point of view, the
linking number $\chi (\gamma _1,\gamma _2)$ may be obtained by introducing
the topological abelian Chern--Simons action \cite{w1}
\begin{equation}
\mathcal{S}_{\mathrm{CS}}(A)=\frac 12\int d^3x\varepsilon ^{\mu \nu \rho
}A_\mu \partial _\nu A_\rho \;,  \label{c-s}
\end{equation}
and by evaluating the correlation function of two loop variables $%
\oint_\gamma dx^\mu A_\mu $, \textit{i..e., }
\begin{equation}
\left\langle \oint_{\gamma _1}dx^\mu A_\mu \oint_{\gamma _2}dy^\nu A_\nu
\right\rangle _{\mathcal{S}_{\mathrm{CS}}}\;.  \label{l-cs}
\end{equation}
That expression (\ref{l-cs}) reproduces the linking number follows from the
observation that the propagator of the gauge field $A_\mu $ obtained from
the Chern--Simons action (\ref{c-s}) upon quantization in the Landau gauge
is precisely the kernel of the Gauss integral (\ref{g-int}), \textit{i.e., }
\begin{equation}
\left\langle A_\mu (x)A_\nu (y)\right\rangle _{\mathcal{S}_{\mathrm{CS}%
}}=\frac 1{4\pi }\varepsilon _{\mu \nu \rho }\frac{(x-y)^\rho }{\left|
x-y\right| ^3}\;.  \label{prop}
\end{equation}
The correlator (\ref{l-cs}) may thus be regarded as a field-theory
description of the linking $\chi (\gamma _1,\gamma _2).$ The action (\ref
{c-s}) can be suitably extended to higher dimensions, providing a
field-theory characterization of the generalizations of the linking number
\cite{tn}. Moreover, the nonabelian version of the three-dimensional
Chern--Simons action (\ref{c-s}) has been proven to play a very relevant
role in knot theory \cite{w1,gu}.

Although the topological field theories possess their own interests and
applications, it is worth underlining here that topological terms appear
frequently as parts of more general effective actions useful for the
theoretical description of a large number of phenomena in different
space-time dimensions. For instance, the effective action corresponding to
the bosonization \cite{bos} of relativistic three-dimensional massive
fermionic systems at $T=0$ can be written as the sum of the Chern-Simons
term (\ref{c-s}) and of an infinite series of higher-order terms in the
curvature $F_{\mu \nu }$ and its derivatives, \textit{i.e., }
\begin{equation}
\mathcal{S}_{\mathrm{CS}}(A)+\mathcal{S}(F),\;  \label{eff-cs}
\end{equation}
with $\mathcal{S}(F)$ being a combination of terms of the type
\begin{equation}
\int d^3xF^2,\;\int d^3xF^4,\;\int d^3xF^6,\;\ldots  \label{comb}
\end{equation}
This kind of action turns out to be useful in order to study several
three-dimensional phenomena such as the Fermi--Bose transmutation \cite
{pk,trasm} and the quantum Hall effect \cite{fr}.

A second interesting example is provided by the five-dimensional
generalization of (\ref{eff-cs}), obtained from the AdS/CFT\textit{\ }%
correspondence \cite{ads,w2}, which relates the conformal $N=4$
super-Yang-Mills theory to type-IIB superstring on AdS$_5\times S^5$. In
fact, in the conformal case, the dual supergravity on $AdS_5$ possesses a
Chern--Simons term obtained from a $SL(2,\Bbb{Z})$ doublet of two-forms $%
B_{\mu \nu }^{RR},$ $B_{\mu \nu }^{NS}$. In this case, the relevant
effective action for $B_{\mu \nu }^{RR},$ $B_{\mu \nu }^{NS}$ looks like
\cite{w2}
\begin{equation}
\int_{\text{AdS}_5}d^5x\varepsilon ^{\mu \nu \rho \lambda \sigma }B_{\mu \nu
}^{RR}\partial _\rho B_{\lambda \sigma }^{NS}\;+\mathcal{S}(dB)\;,
\label{ads}
\end{equation}
where the term $\mathcal{S}(dB)$ collects all the higher-order terms in the
curvatures $dB^{RR},dB^{NS}$. The correlation function (\ref{l-cs})
generalizes now to
\begin{equation}
\left\langle \int_{\Sigma _1}d\sigma ^{\mu \nu }B_{\mu \nu
}^{RR}\int_{\Sigma _2}d\sigma ^{\lambda \rho }B_{\lambda \rho
}^{NS}\right\rangle \;,  \label{gen}
\end{equation}
where $\Sigma _1,$ $\Sigma _2$ are appropriate two-surfaces.

In view of these applications, it seems natural to ask ourselves what is the
response of a correlator of the type (\ref{l-cs}) when the corresponding
topological field theory is perturbed by the introduction of a
nontopological interaction term depending on the curvature.

This is the aim of the present paper. More precisely, we shall report on the
four-loop computation of the correlator
\begin{equation}
\left\langle \oint_{\gamma _1}dx^\mu A_\mu \oint_{\gamma _2}dy^\nu A_\nu
\right\rangle _{\mathcal{S}_{\mathrm{eff}}}\;,  \label{p-c}
\end{equation}
when the three-dimensional Chern--Simons action (\ref{c-s}) is perturbed by
a nontopological interaction term of the kind $\int F^4,$ namely\textit{, }%
expression (\ref{p-c}) will be evaluated with an effective action $\mathcal{S%
}_{\mathrm{eff}}$ given by
\begin{equation}
\mathcal{S}_{\mathrm{eff}}=\frac 12\int d^3x\varepsilon ^{\mu \nu \rho
}A_\mu \partial _\nu A_\rho +\frac \tau {4!}\int d^3x\widetilde{F}^\mu
\widetilde{F}_\mu \widetilde{F}^\nu \widetilde{F}_\nu \;,  \label{seff}
\end{equation}
with $\widetilde{F}^\mu =\frac 12\varepsilon ^{\mu \nu \rho }F_{\nu \rho }$
and $\tau $ being an arbitrary parameter with negative mass dimension,
reflecting the power-counting nonrenormalizability of the perturbation.

In particular, we shall be able to prove that the correlation function (\ref
{p-c}) turns out to be independent from $\tau $, yielding the linking number
$\chi (\gamma _1,\gamma _2)$ of the two curves $\gamma _1$, $\gamma _2.$
Although the loop analysis will be worked out only up to the fourth order,
this conclusion holds to all orders of perturbation theory and may be easily
generalized to any local nontopological interaction term containing
arbitrary powers of the curvature $F_{\mu \nu }$ as well as to the
higher-dimensional cases \cite{pr} as, for instance, the effective action of eq.(\ref
{ads}).

This result means that the loop correlator (\ref{p-c}) is stable with
respect to the perturbations which can be added to the starting topological
action. In other words, the expression (\ref{p-c}) will give the linking
number of the two curves $\gamma _1$, $\gamma _2$, regardless of any $F$%
-dependent perturbation term that can be introduced and of their
power-counting nonrenormalizability character.

Two remarks are now in order. First, we will limit here ourselves only to
effective actions which are abelian. Second, we shall consider only $F$%
-dependent terms which can be treated as true perturbations. Therefore, we
shall avoid in the effective action (\ref{seff}) the inclusion of a term of
the Maxwell type
\begin{equation}
\mathcal{S}_{\mathrm{Max}}=\frac 1{4m}\int d^3xF^{\mu \nu }F_{\mu \nu }\;,
\label{max}
\end{equation}
where $m$ is a mass parameter. The presence of this term would completely
modify the original properties of the model. In fact, being expression (\ref
{max}) quadratic in the gauge fields, it cannot be considered as a
perturbation term, as it will be responsible for the presence of massive
excitations in the spectrum of the theory \cite{mcs}. Rather, the presence
of the Maxwell term in the effective action (\ref{seff}) will give rise to
the existence of two distinct regimes corresponding to the long and short
distance behaviours, respectively. For distances larger than the inverse of
the mass parameter $m$ (\textit{i.e.}, the low-energy regime ), the
topological term will prevail,\textit{\ }while the Maxwell term will become
the relevant one at short distances (\textit{i.e.,} the high-energy regime).
It is worth mentioning here that these two regimes can be accessed in a very
simple way by means of suitable gauge-invariant field redefinitions of the
gauge connection $A_\mu $ \cite{largem}. However, their full understanding
is a difficult and delicate task, which is beyond the aim of the present paper,
being under investigation.

We should also underline here that, in the abelian case, the loop variable $%
\oint_\gamma dx^\mu A_\mu $ is gauge invariant for closed curves, and so
there is no need to take into account its exponentiation $e^{i\oint_\gamma
dx^\mu A_\mu }$, as it would be required in the nonabelian case. This
feature has a useful consequence. It allows indeed to avoid the case in
which the double-line integral (\ref{p-c}) has to be taken along the same
curve. This case, usually referred to as  the self-linking, would be
automatically generated by the perturbative Taylor expansion of the
exponential $e^{i\oint_\gamma dx^\mu A_\mu }$. In other words, as far as the
abelian case is concerned, the loop variables in eq.(\ref{p-c}) do not need
to be exponentiated. Therefore, the two curves $\gamma _1$ and $\gamma _2$
will always refer to two distinct curves which do not intersect each other.
As we shall see in the following, this point will be relevant in order to
establish the independence from the parameter $\tau $ of the expression (\ref
{p-c}).

\section{Perturbative expansion and Feynman diagrams}

In order to discuss the perturbative expansion of the loop correlator (\ref
{p-c}), let us first define the gauge-fixed version of the effective action
which shall be used throughout the present article, namely,
\begin{equation}
\mathcal{S}_{\mathrm{eff}}=\frac 12\int d^3x\varepsilon ^{\mu \nu \rho
}A_\mu \partial _\nu A_\rho +\int d^3x\,b\partial A+\frac \tau {4!}\int d^3x:%
\widetilde{F}^\mu \widetilde{F}_\mu \widetilde{F}^\nu \widetilde{F}_\nu :\,,
\label{effective}
\end{equation}
where the Lagrange multiplier $b$ has been introduced in order to implement
the Landau gauge. Notice that we have Wick-ordered the quartic interaction
term, which will allow to rule out tadpole diagrams.\footnote{%
We remind the reader that, in the present abelian case, the normal-ordering
prescription is compatible with the requirement of gauge invariance. This
follows from the observation that the positive and negative-frequency parts $%
F_{\mu \nu }^{(+)}$ and $F_{\mu \nu }^{(-)}$ of the field strength $F_{\mu
\nu }$ are each gauge invariant.}

As usual in this kind of problem, we shall make use of the configuration
space rather than the momentum space. Let us now give the elementary Wick
contractions which shall be needed for the evaluation of the Feynman
diagrams. Recalling that
\begin{equation}
\partial ^2\frac 1{\left| x-y\right| }=-4\pi \delta ^3(x-y)\;,  \label{inv-l}
\end{equation}
from eq.(\ref{prop}) one obtains
\begin{equation}
\left\langle A_\mu (x)\widetilde{F}_\nu (y)\right\rangle =g_{\mu \nu }\delta
^3(x-y)+\partial _\mu \partial _\nu \frac 1{4\pi \left| x-y\right| }\;,
\label{wc1}
\end{equation}
and
\begin{equation}
\left\langle \widetilde{F}_\mu (x)\widetilde{F}_\nu (y)\right\rangle
=-\varepsilon _{\mu \nu \rho }\partial ^\rho \delta ^3(x-y)\;.  \label{wc2}
\end{equation}
Concerning now the perturbative loop expansion, it is easily checked that
the first Feynman diagram which contributes to the correlation function (\ref
{p-c}) is of two-loop order and can be drawn as follows:

\begin{center}
\begin{figure}[h]
\setlength{\unitlength}{1mm}
\hspace{2cm}\scalebox{0.75}{\includegraphics*{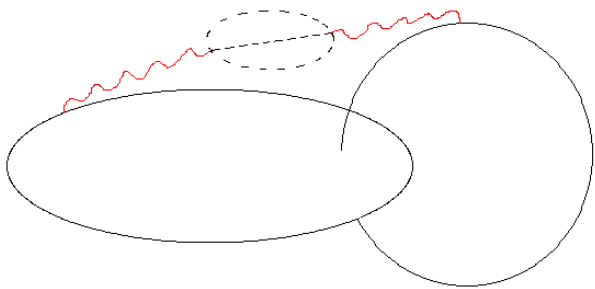}}
\caption{Two-loop contribution}
\end{figure}
\end{center}


In the above figure, the wavy and dashed lines refer
respectively to the  Wick contractions $\left\langle
A\widetilde{F}\right\rangle$ and
 $\left\langle\widetilde{F}\widetilde{F}\right\rangle$.

The Feynman integrals corresponding to the diagram of Fig.2
are easily written down by means of eqs.(\ref{wc1}),
(\ref{wc2}). However, before computing them, let us spend a
few words on the mechanism which is responsible for the
independence on the parameter $\tau $ of expression (\ref
{p-c}). From the structure of the diagram of Fig.2, we
observe that the gauge fields $A_\mu (x)$ and $A_\nu (y)$
lying on the two curves $\gamma
_1\; $and $\gamma _{2\text{ }}$will be always contracted with the $%
\widetilde{F}$'s present in the interaction term of the expression (\ref
{effective}). Therefore, besides contractions of the type $\left\langle
\widetilde{F}\widetilde{F}\right\rangle $, the corresponding Feynman
integrals will always contain two contractions of the kind $\left\langle A%
\widetilde{F}\right\rangle $. However, one should remark in the second term
of eq.(\ref{wc1}) that one of the Lorentz indices of the two space-time
derivatives corresponds to the vector index of a gauge field lying on either
$\gamma _1\; $or $\gamma _2.$ It thus refers to a total derivative with
respect to the variable running along one of the closed loops, implying a
vanishing contribution. In other words, the second term of eq.(\ref{wc1})
may be neglected. As a consequence, all the Wick contractions entering the
Feynman integrals will basically lead to a product of delta functions. After
the introduction of a suitable regularization, the latter can be integrated
out, finally resulting in a $\delta ^3(x-y)$, where, we remind, $x$ and $y$
run along each of the two curves, respectively. As these variables never
coincide, the whole expression vanishes identically, ensuring the
independence from the parameter $\tau $ of the correlator (\ref{p-c}). The
same mechanism can be seen to occur at higher loop orders, as it will be
explicitly shown later on.

Apart from an irrelevant global symmetry coefficient, the
diagram of Fig.2 therefore corresponds to the following
integral:
\begin{eqnarray}
I^{(2)}\!\!\! &=&\!\!\!\oint_{\gamma _1}dx^\mu \oint_{\gamma _2}dy^\nu \int
d^3z_1\,d^3z_2\,\left\langle A_\mu (x)\tilde{F}_\alpha (z_1)\right\rangle
\left\langle A_\nu (y)\tilde{F}_\gamma (z_2)\right\rangle \left\langle
\tilde{F}_\beta (z_1)\tilde{F}_\rho (z_2)\right\rangle  \nonumber \\
&&\!\!\!\!\!\!\times \left[ 4\left\langle \tilde{F}^\alpha (z_1)\tilde{F}^\rho
(z_2)\right\rangle \left\langle \tilde{F}^\beta (z_1)\tilde{F}^\gamma
(z_2)\right\rangle
+ 2\left\langle \tilde{F}^\alpha (z_1)\tilde{F}%
^\gamma (z_2)\right\rangle \left\langle \tilde{F}^\beta (z_1)\tilde{F}^\rho
(z_2)\right\rangle
 \right] .  \nonumber \\
&&\,\,  \label{int2}
\end{eqnarray}
Let us analyse the first term of the above expression. Making use of the
propagators (\ref{wc1}) and (\ref{wc2}), we obtain
\begin{eqnarray}
&&-4\oint_{\gamma _1}dx^\mu \oint_{\gamma _2}dy^\nu \int d^3z_1\,d^3z_2\,
\nonumber \\
&&\quad \times \left[ g_{\mu \alpha }\delta ^3(x-z_1)+\partial _\mu \partial
_\alpha \frac 1{4\pi \left| x-z_1\right| }\right] \left[ g_{\nu \gamma
}\delta ^3(y-z_2)+\partial _\nu \partial _\gamma \frac 1{4\pi \left|
y-z_2\right| }\right]  \nonumber \\
&&\quad \times \left[ \varepsilon _{\beta \rho \lambda }\partial ^\lambda
\delta ^3(z_1-z_2)\right] \left[ \varepsilon ^{\alpha \rho \tau }\partial
_\tau \delta ^3(z_1-z_2)\right] \left[ \varepsilon ^{\beta \gamma \sigma
}\partial _\sigma \delta ^3(z_1-z_2)\right] \,.  \label{second}
\end{eqnarray}
As previously mentioned, the terms containing the derivatives $\partial _\mu
$ and $\partial _\nu $ do not contribute, as they correspond to total
derivatives on closed curves. Expression (\ref{second}) then becomes
\begin{eqnarray}
&&4\oint_{\gamma _1}dx^\mu \oint_{\gamma _2}dy^\nu \int
d^3z_1\,d^3z_2\,\delta ^3(x-z_1)\delta ^3(y-z_2)  \nonumber \\
&&\quad \times \left[ \varepsilon _{\beta \rho \lambda }\partial ^\lambda
\delta ^3(z_1-z_2)\right] \left[ \varepsilon _\mu {}^{\rho \tau }\partial
_\tau \delta ^3(z_1-z_2)\right] \left[ \varepsilon ^{\beta \sigma }{}_\nu
\partial _\sigma \delta ^3(z_1-z_2)\right] \,.  \label{second2}
\end{eqnarray}
In spite of the presence of products of delta functions with the same
arguments, the above expression is easily seen to vanish. Let us show this
claim in two ways. First, we observe that there is always a possible order
of taking the integrations over the delta functions such that we end up with
products of $\delta ^3(x-y)$ and not of $\delta ^3(0)$. In the present case,
this would amount to integrate out first the two delta functions with
arguments $x-z_1$ and $y-z_2$, which would lead to
\begin{equation}
4\oint_{\gamma _1}dx^\mu \oint_{\gamma _2}dy^\nu \left[ \varepsilon _{\beta
\rho \lambda }\partial ^\lambda \delta ^3(x-y)\right] \left[ \varepsilon
_\mu {}^{\rho \tau }\partial _\tau \delta ^3(x-y)\right] \left[ \varepsilon
^{\beta \sigma }{}_\nu \partial _\sigma \delta ^3(x-y)\right] =0\,,
\label{second3}
\end{equation}
since $x-y$ never vanishes. It is worth remarking that this possibility exists, in fact, for
the higher-order diagrams, as will be shown below.

Second, we can adopt a more rigorous treatment by regularizing the delta
functions with coinciding arguments through the point-splitting procedure
already used by Polyakov \cite{pk}:
\begin{eqnarray}
\delta _\varepsilon (z_1-z_2) &=&\frac 1{\left( 2\pi \varepsilon \right)
^{3/2}}e^{-(z_1-z_2)^2/2\varepsilon }\,,  \label{delreg} \\
\lim_{\varepsilon \rightarrow 0}\delta _\varepsilon (z_1-z_2) &=&\delta
^3(z_1-z_2)\;.  \nonumber
\end{eqnarray}
More precisely, whenever a product of $n$ delta functions with coinciding arguments
occurs, it
will be understood  as
\[
\left[ \delta ^3(z_1-z_2)\right] ^n=\left[ \delta _\varepsilon
(z_1-z_2)\right] ^{n-1}\delta ^3(z_1-z_2)\,,
\]
where the limit $\varepsilon \rightarrow 0$ is meant to be taken at the end
of all calculations. Accordingly, expression (\ref{second2}) will be
replaced by its regularized version,
\begin{eqnarray}
&&\lim_{\varepsilon \rightarrow 0}4\oint_{\gamma _1}dx^\mu \oint_{\gamma
_2}dy^\nu \,\delta ^3(x-z_1)\delta ^3(y-z_2)  \nonumber \\
&&\quad \times \left[ \varepsilon _{\beta \rho \lambda }\partial ^\lambda
\delta _\varepsilon (z_1-z_2)\right] \left[ \varepsilon _\mu {}^{\rho \tau
}\partial _\tau \delta _\varepsilon (z_1-z_2)\right] \left[ \varepsilon
^{\beta \sigma }{}_\nu \partial _\sigma \delta ^3(z_1-z_2)\right] \,.
\label{second4}
\end{eqnarray}
Whatever the order of integration, we get, before taking the limit, an
expression containing $\delta ^3(x-y)$, which leads to a null result.

The second term of (\ref{int2}) follows analogously, so that the two-loop
diagram of Fig.2 does not contribute to the correlator (\ref{p-c}).

Concerning the higher-order contributions in the
perturbation theory, the results are of a similar nature.
The topologically distinct diagrams contributing to the 3-
and 4-loop are given in Fig.3 and in Figs.4 and 5.

\begin{center}
\begin{figure}
\setlength{\unitlength}{1mm}
\hspace{2cm}\scalebox{0.75}{\includegraphics*{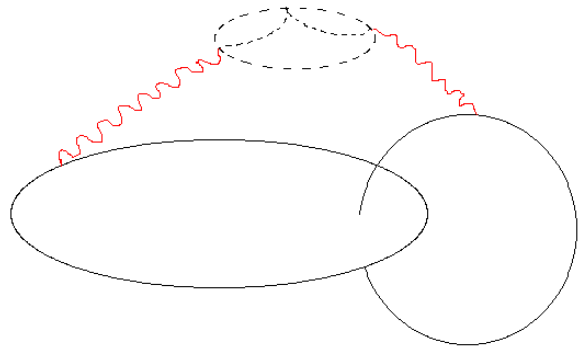}}
\caption{Three-loop contribution}
\end{figure}
\end{center}


\begin{center}
\begin{figure}
\setlength{\unitlength}{1mm}
\hspace{2cm}\scalebox{0.75}{\includegraphics*{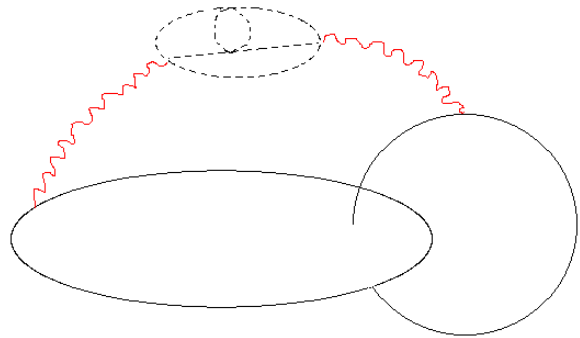}}
\caption{Four-loop contribution, first type}
\end{figure}
\end{center}


\newpage

\begin{center}
\begin{figure}
\setlength{\unitlength}{1mm}
\hspace{2cm}\scalebox{0.75}{\includegraphics*{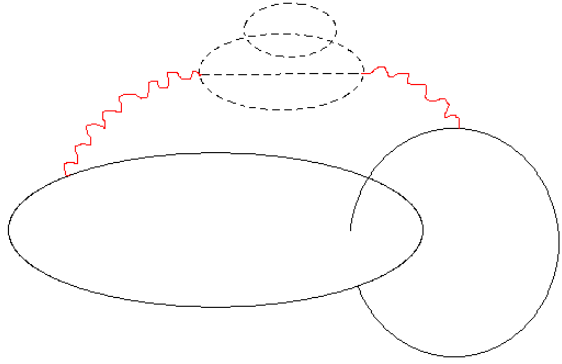}}
\caption{Four-loop contribution, second type}
\end{figure}
\end{center}


It is sufficient to present here just one typical term of each order. A
notational simplification is convenient. We define the transverse derivative
operator
\[
\tilde{\partial}_{\mu \nu }\equiv \varepsilon _{\mu \nu \rho }\partial ^\rho
.
\]
For instance, a typical contraction from Fig.3 is
proportional to
\begin{eqnarray}
I^{(3)} &=&\oint_{\gamma _1}dx^\mu \oint_{\gamma _2}dy^\nu \int
d^3z_1\,d^3z_2d^3z_3\,g_{\mu \alpha }\delta ^3(x-z_1)g_{\nu \gamma }\delta
^3(y-z_2)  \nonumber \\
&&\times \left[ \tilde{\partial}^{\alpha \gamma }\delta ^3(z_1-z_2)\right]
\left[ \tilde{\partial}_{\beta \delta }\delta ^3(z_1-z_3)\right] \left[
\tilde{\partial}^{\beta \lambda }\delta ^3(z_1-z_3)\right]  \nonumber \\
&&\times \left[ \tilde{\partial}_{\rho \lambda }\delta ^3(z_2-z_3)\right]
\left[ \tilde{\partial}^{\rho \delta }\delta ^3(z_2-z_3)\right]  \nonumber \\
&=&\oint_{\gamma _1}dx^\mu \oint_{\gamma _2}dy^\nu \int d^3z_3\left[ \tilde{%
\partial}_{\mu \nu }\delta ^3(x-y)\right]  \nonumber \\
&&\times \left[ \tilde{\partial}_{\beta \delta }\delta _\varepsilon
(x-z_3)\right] \left[ \tilde{\partial}^{\beta \lambda }\delta
^3(x-z_3)\right] \left[ \tilde{\partial}_{\rho \lambda }\delta _\varepsilon
(y-z_3)\right] \left[ \tilde{\partial}^{\rho \delta }\delta
^3(y-z_3)\right] ,  \nonumber \\
&&  \label{third}
\end{eqnarray}
while the diagram of Fig.4 gives
\begin{eqnarray}
I^{(4)} &=&\oint_{\gamma _1}dx^\mu \oint_{\gamma
_2}dy^\nu \int d^3z_1\cdots d^3z_4\,g_{\mu \alpha }\delta
^3(x-z_1)g_{\nu \gamma }\delta
^3(y-z_2)  \nonumber \\
&&\times \left[ \tilde{\partial}^{\alpha \gamma }\delta ^3(z_1-z_2)\right]
\left[ \tilde{\partial}^{\beta \lambda }\delta ^3(z_1-z_3)\right] \left[
\tilde{\partial}_{\beta \sigma }\delta ^3(z_1-z_4)\right]  \nonumber \\
&&\times \left[ \tilde{\partial}^{\rho \sigma }\delta ^3(z_2-z_4)\right]
\left[ \tilde{\partial}_{\rho \lambda }\delta ^3(z_2-z_3)\right] \left[
\tilde{\partial}^{\varphi \omega }\delta ^3(z_3-z_4)\right]  \nonumber \\
&&\times \left[ \tilde{\partial}_{\varphi \omega }\delta ^3(z_3-z_4)\right]
\nonumber \\
&=&\oint_{\gamma _1}dx^\mu \oint_{\gamma _2}dy^\nu \int
d^3z_3\,d^3z_4\,\left[ \tilde{\partial}_{\mu \nu }\delta ^3(x-y)\right]
\nonumber \\
&&\times \left[ \tilde{\partial}^{\beta \lambda }\delta ^3(x-z_3)\right]
\left[ \tilde{\partial}_{\beta \sigma }\delta ^3(x-z_4)\right] \left[ \tilde{%
\partial}^{\rho \sigma }\delta ^3(y-z_4)\right]  \nonumber \\
&&\times \left[ \tilde{\partial}_{\rho \lambda }\delta ^3(y-z_3)\right]
\left[ \tilde{\partial}^{\varphi \omega }\delta _\varepsilon
(z_3-z_4)\right] \left[ \tilde{\partial}_{\varphi \omega }\delta^3(z_3-z_4)\right] .
\label{fourth}
\end{eqnarray}
In order to obtain the above expressions we have followed the same
prescription established before for regularizing the delta functions with
identical arguments. Notice also that we have integrated out first the two
$\delta$-functions whose arguments depend on the points $x$ and $y$ of the two curves.

All terms in all possible diagrams may then be seen to be proportional to $%
\delta ^3(x-y)$ (or its derivatives). One may easily convince oneself that
this mechanism also applies to any order in perturbation theory. As we
always have $x\neq y$, these diagrams all amount to a null correction to the
basic diagram, so that the correlation function (\ref{p-c}) for two closed
smooth nonintersecting curves $\gamma _1,\gamma _2$ gives their linking
number to all orders:
\begin{equation}
\left\langle A_\mu (x)A_\nu (y)\right\rangle _{\mathcal{S}_{\text{eff }%
}}=\chi (\gamma _1,\gamma _2).  \label{finalres}
\end{equation}

\section{Conclusions}

We have been able to show, in the present article, that the correlation
function (\ref{p-c}) is unaffected by the radiative corrections, provided $%
\gamma _1,\gamma _2$ are two nonintersecting closed curves. Although we have
given explicit expressions for the $\int \tilde{F}^4$ perturbation, the same
result may be achieved for any local interaction term of the type $\int
\tilde{F}^n$.

We may interpret this result as a kind of nonrenormalization property of the
linking number, reflecting its stability with respect to any local gauge
invariant perturbation of the starting Chern-Simons action.

Further generalizations to higher dimensions as well as to the nonabelian
case are under investigation \cite{pr}.

\vspace{2cm}

{\Large \textbf{Acknowledgements}}

The Conselho\ Nacional de Pesquisa e Desenvolvimento (CNPq/Brazil), the
Funda\c {c}\~{a}o de Amparo \`{a} Pesquisa do Estado do Rio de Janeiro
(Faperj) and the SR2-UERJ are gratefully acknowledged for financial support.

\end{document}